\documentclass[aps,a4paper,twocolumn,showpacs,showkeys,preprintnumbers,amsmath,amssymb]{revtex4}
\usepackage{graphicx}
\usepackage{bm}
\usepackage{dcolumn}
\usepackage{color}
\usepackage{pst-plot}
\renewcommand{\texttt}{{}}
\newcommand{\be}{\begin{eqnarray}}
\newcommand{\ee}{\end{eqnarray}}
\newcommand{\nn}{\nonumber}
\usepackage{graphicx}

\begin{document}

%discontinuity 

%\title{The microstructure of a quantum spacetime}
%\title{Kerr and Kerr-Newman black holes in the presence of a minimal lenght} % of Loop Quantum Gravity}
%\title{Non-commutative Kerr-Newmann black hole}% in presence of a minimal length}
%\title{Spinning and charged 
%\title{Regular Axisymmetric Black Holes} % in presence of a minimal length}
%\title{KerrR and KerrR-Newman black holes via Newman-Janis algorithm} 
%\title{Axisymmetric black holes in presence of a minimal length}
%\title{Noncommutative geometry inspired spinning charged black holes via Newman-Janis algorithm}
%\title{Noncommutative geometry inspired Kerr and Kerr-Newman black holes}
%\title{Noncommutative %geometry 
%inspired rotating and charged
%higher dimensional black holes}
%Spinning Noncommutative black holes in higher dimensional space-times
%\title{Rotating Noncommutative Black Holes} % in higher dimensional space-times}
%\title{Charged rotating regular black holes} % in higher dimensional space-times}\\title
\title{Charged rotating noncommutative black holes}

\author{Leonardo Modesto}
\thanks{lmodesto@perimeterinstitute.ca}
\affiliation{Perimeter Institute for Theoretical Physics, 31 Caroline Street, Waterloo, Ontario N2L 2Y5, Canada}

\author{Piero Nicolini}
\thanks{nicolini@th.physik.uni-frankfurt.de}
\affiliation{Frankfurt Institute for Advanced Studies (FIAS),
%Johann Wolfgang Goethe University
Institut f\"ur Theoretische Physik, Johann Wolfgang Goethe-Universit\"at, 
Ruth-Moufang-Strasse 1, 60438 Frankfurt am Main, Germany}

\date{\small\today}

\begin{abstract} In this paper we complete the program of the noncomutative geometry inspired black holes, providing the richest possible solution, endowed with mass, charge and angular momentum. After providing a prescription for employing the Newman-Janis algorithm in the case of nonvanishing stress tensors, we find regular axisymmetric charged black holes in the presence of a minimal length. We study also the new thermodynamics and we determine the corresponding higher-dimensional solutions. As a conclusion we make some consideration about possible applications.

\end{abstract}
%\pacs{05.45.Df, 04.60.Pp}

\maketitle

\tableofcontents

\section{Introduction}

When we talk about curvature singularities in general relativity, we often forget that they are just fictitious effects emerging where the classical description of the gravitational field breaks down. In other words, to obtain physically reliable scenarios in spacetime regions plagued by curvature singularities, one has just the only possibility of invoking some formulation of the quantum theory of gravitation. Along this line of reasoning, new black hole geometries have successfully been derived, by improving the classical singularities with quantum gravity mechanisms. For instance, loop quantum black holes (LQBHs) have been obtained starting from the prescriptions of  loop quantum qravity. Inside the event horizon, a quantized version of the Kantowski-Sachs spacetime has replaced the irregular Schwarzschild geometry \cite{LQBHs}. Other examples are the noncommutative geometry inspired black holes (NCBHs), whose line element singularities are cured by the presence of a minimal length effectively induced by the noncommutative character of coordinate operators \cite{review,Nicolini:2005vd,NCBHs,NCBHs2}. Even if LQBHs and NCBHs use different tools to tame the curvature singularity, they share many common features, which would suggest an universal character when quantum gravity effects are taken into account. For instance, under some additional hypotheses, they both admit a maximum temperature, showing a thermodynamically stable final phase of the Hawking evaporation, or they both can be dirty, namely admitting  $g_{00}$ metric components which do not coincide with the usual value $-g_{11}^{-1}$ \cite{dirty}.
While LQBHs are for now known only for neutral, nonrotating, static four-dimensional geometries, the program of NCBHs includes a richer variety of physical situations: indeed the charged solution \cite{Ansoldi:2006vg}, the extradimensional neutral \cite{rizzo} and charged solutions \cite{extracharged} and the rotating one \cite{kerrr} have been obtained and they all exhibit a  deSitter core/belt as a consequence of quantum  manifold fluctuations at the origin. The general strategy to derive NCBH solutions consists of prescribing an improved form of the energy-momentum tensor, which accounts for the noncommutative fluctuations of the manifold at the origin and being vanishing for distances larger with respect to the noncommutative geometry typical scale, i.e. the minimal length. Another requirement is the covariant conservation of the energy-momentum tensor, while energy conditions at the origin are generally violated, in agreement with the fact that we are invoking a non-standard, i.e. non-classical, kind of matter. The derivation of the correct form of the energy-momentum tensor requires great care, in particular, for the axisymmetric case. Indeed, thanks to some considerations about the Kerr-Schild decomposition and some physical requirements for the regularity of the manifold, the noncommutative inspired Kerr (NCKerr) solution has been derived starting from a spinning fluid-type energy-momentum tensor. On the other hand, even for the spherically symmetric case, the inclusion of the electromagnetic interaction, makes the determination of the energy-momentum tensor much more difficult. In particular, if we want to have a complete scenario for a regular black hole endowed with all the three externally observable classical  parameters, mass, electric charge, and angular momentum, we need to devise an alternative to the direct derivation of the energy-momentum tensor.
In this paper, we propose a way to circumvent the aforementioned difficulties opening a new route on the basis of the Newman-Janis algorithm \cite{NJA}. However this kind of algorithm works only for vacuum or Reissner-Nordstr\"{o}m geometries. As a result, we first provide a prescription to amend the Newman-Janis algorithm and extend its validity to the case of NCBHs. Then we show how this prescription lets us correctly obtain the recent regular spinning black hole, i.e. the NCKerr solution. Then we proceed by deriving the corresponding regular Kerr-Newman (NCKN) solution and the higher-dimensional version of NCKerr and NCKN metrics.
We recall that there exist other attempts to determine regular spinning black hole solutions \cite{otherkerrr}. However the strategy at the basis of these derivations has been always that of matching an outer geometry with a regular local geometry at the origin. Conversely, by means of the Newman-Janis mechanism we shall present unique solutions which interpolate the local regular behavior at the origin and the conventional rotating geometries at large distances. In addition, against the existing literature, our solutions have an intrinsic quantum gravitational character at the origin, since the Newman-Janis algorithm maintains the original features of the corresponding nonrotating NCBHs. 

The paper is organized as follows.  In Sec. \ref{NJA}, we present the Newman-Janis algorithm which we will use in the following sections to derive the NCKerr solution (Sec. \ref{kerrr}), the NCKN (Sec. \ref{kerrrn}) and the higher-dimensional NCKerr and NCKN solutions (Sec. \ref{extrakerrr}). Throughout the sections, discussions are devoted both to regularity of the geometry and the behavior of the Hawking temperature. In Sec. \ref{concl}, we draw the conclusions.

%%\textbf{\begin{itemize}
%%\item \textbf{black hole electron}
%%\item \textbf{geons in quantum gravity}
%%\item \textbf{extremal black hole}
%%\end{itemize}}

\section{Newman-Janis algorithm}
\label{NJA}
The Newman-Janis algorithm is often regarded as a short cut to obtain spinning black hole solutions from the corresponding nonrotating ones. According to \cite{Drake:1998gf}, the algorithm works only for vacuum solutions or the case of a Maxwell stress tensor at the most. After presenting the general procedure, we will provide in the next section a prescription to include the case of nonvanishing stress tensors.
We start from the line element  
\be
&& ds^2=e^{2\Phi(r)}dt^2-e^{2\lambda(r)}dr^2 - H(r)  d\Omega^2 \nonumber \\
&& \hspace{0.6cm} = G(r) dt^2-\frac{dr^2}{F(r)}-H(r) d\Omega^2.
\label{GSS}
\ee
Employing the outgoing  Eddington-Finkelstein coordinates $\left\{u,r,\vartheta, \phi\right\}$, where 
\be
u=t-r^\ast
\ee
and $dr^\ast= dr/\sqrt{G F}$, the above line element can be written in the following shape
\be
ds^2 = G(r) du^2 + 2 \sqrt{\frac{G(r)}{F(r)}} \,  du dr - H(r) \, d\Omega^2.
\nonumber
\ee
The covariant metric in matrix form reads 
\begin{eqnarray}
g^{\mu \nu} = \left(
\begin{array}{cccc}
0 & e^{- \Phi(r) - \lambda(r)} & 0 & 0 \\
 e^{- \Phi(r) - \lambda(r)} & - e^{- 2 \lambda(r)} & 0 & 0 \\
 0 & 0 &  \frac{- 1}{H(r)}  & 0 \\
 0 & 0 & 0& \frac{ - 1}{H(r) \sin^2 \vartheta}
\end{array}
\right) \hspace{-0.1cm}. \end{eqnarray}
We can write the metric in terms of null tetrad vectors
\be
&& g^{\mu \nu } = l^{\mu} n^{\nu} +  l^{\nu} n^{\mu} -  m^{\mu} {\bar m}^{\nu} -  m^{\nu} {\bar m}^{\mu} ,\nonumber\\
&& l^{\mu } = \delta_1^{\mu} ,  \nonumber \\
&& n^{\mu} = \sqrt{\frac{F}{G}} \delta^{\mu}_0 - \frac{1}{2} F \delta^{\mu}_1, \nonumber \\
&& m^{\mu} = \frac{1}{\sqrt{ 2 H}} \left( \delta_2^{\mu} + \frac{ i}{\sin \vartheta} \delta_3^{\mu} \right)
\label{tetra}
\ee
where 
$l_{\mu} l^{\mu} = m_{\mu} m^{\mu} = n_{\mu} n^{\mu} = l_{\mu} m^{\mu} =n_{\mu} m^{\mu} =0$
and $l_{\mu} n^{\mu} = - m_{\mu} {\bar m}^{\mu} =1$ (${\bar x}$ is the complex conjugate of the general quantity $x$).
Complex null tetrads form the starting point to derive the {\em Kerr spinning black hole metric}. 
To this purpose, we consider the following {\em Newman-Janis} complex increment 
\begin{eqnarray}
&& r \rightarrow r' = r + i \, a \cos \vartheta , \nonumber \\
&&  u \rightarrow u' = u  - i \, a \cos \vartheta.
\label{NJ}
\end{eqnarray}
%where in general we assume also $a(r)$ to be a function of the radial coordinate.
Under the complex increment the tetrads change 
\be
&& l^{\mu } = \delta_1^{\mu}, \\ %\nonumber \\
&& n^{\mu} = \sqrt{\frac{F(r, \vartheta)}{G(r, \vartheta)}} \, \delta^{\mu}_0 - \frac{1}{2} F(r, \vartheta)\,  \delta^{\mu}_1, \nonumber \\
&& m^{\mu} = \frac{1}{\sqrt{ 2 H(r, \vartheta)}} 
\left(i a \sin \vartheta (\delta^{\mu}_0 - \delta^{\mu}_1) + \delta_2^{\mu} + \frac{ i}{\sin \vartheta} \delta_3^{\mu} \right). \nonumber 
\label{tetra2}
\ee
%\textbf{aggiungere la formula generale in forma matriciale
%}
%%%%%%%%%%

For $G =F$ and $H= r^2$ the metric simplifies and reads 
\be
ds^2 = G(r) du^2 + 2dudr -r^2 d\Omega^2.
\nonumber
\ee
\section{Regular Kerr black hole}
\label{kerrr}
\subsection{Derivation of noncommutative black holes}

We start by briefly recalling the general ideas behind NCBH solutions. The quest for a noncommutative formulation of general relativity is a long standing problem which seems, up to now far from being close to a solution. The general procedure of employing the Moyal $\star$-product among vielbein fields in gravity action is mathematically correct but physically not efficient when studying specific solutions. This can be explained by the fact that, for practical computations one must expand the $\star$-product in the noncommutative parameter: at any truncation at a desired order one actually destroys the nonlocal character of the theory \cite{Chamseddine:1992yx}. As a consequence, the resulting black hole geometries are affected by a singular behavior at the origin as conventional classical solutions \cite{Chaichian:2007we}.
Against this background, there is a possibility of circumventing the problem by changing strategy. Instead of formulating a theory with a nonlocal product one may think to implement the nonlocal character of noncommutative geometry, by an effective deformation of conventional field equations. Roughly speaking one can average noncommutative fluctuations of the manifold and work with the resulting mean values. The starting point is the ultimate fate of the classical point-like object in noncommutative geometry. In a series of papers based on the coordinates coherent state approach to noncommutative geometry \cite{NCQFT} it has been shown that the mean position of a point like object in a noncommutative manifold is no longer governed by a Dirac delta function but by a Gaussian distribution
\begin{equation}
\rho_{\theta}(\vec{x})=\frac{1}{(4\pi\theta)^{d/2}} e^{-\vec{x}^2/4\theta},
\end{equation}
where $d$ is the manifold dimension and $\theta$ the noncommutative parameter with dimensions of a length squared that encodes a minimal length in the manifold. In addition it has been shown that primary corrections to any field equation in the presence of a noncommutative background can be obtained by replacing the conventional point-like source term (matter sector) with a Gaussian distribution, while keeping formally unchanged differential operators (geometry sector) \cite{NCBHs}. In the specific case of the gravity field equations this is equivalent to saying that the only modification occurs at the level of the energy-momentum tensor,  while $G_{\mu\nu}$ is formally left
unchanged.

For a static, spherically symmetric, noncommutative diffused, particlelike gravitational source, one gets a Gaussian profile for the $T_0^0$ component of the  energy-momentum tensor. The covariant conservation and the additional ``Schwarzschild-like'' condition $g_{00}=-g_{rr}^{-1}$ completely specify the form of the energy-momentum tensor which generates the solution \cite{Nicolini:2005vd} 
\be
ds^2= 
 \left(1-\frac{2m(r)}{r}\right)dt^2-\frac{dr^2}{\left(1-\frac{2m(r)}{r}\right)}-d\Omega^2 ,
 \label{ncschw}
\ee
where
\be
m(r)=M\ \frac{\gamma(3/2;\ r^2/4\theta)}{\Gamma(3/2)},
\ee
with $M$ the total (constant) mass of the system and
\be
&& \gamma(3/2;\ r^2/4\theta)=\int_0^{r^2/4\theta}t^{1/2}e^{-t}dt  \, , \nonumber \\   %\,\,\,\,\,\, 
&& \Gamma(3/2)=\int_0^{\infty}t^{1/2}e^{-t}dt=\frac{\sqrt{\pi}}{2}. 
\ee
This metric describes the noncommutative geometry inspired Schwarzschild (NCSchw) black hole. At large distances, i.e. $r\gg\sqrt\theta$, the function $m(r)$ approaches the total mass $M$, namely $m(r)\to M$, and 
(\ref{ncschw}) matches the conventional Schwarzschild solution. At short distances $r\ll\sqrt\theta$ a regular de Sitter core accounts for the manifold fluctuations, taming the singularity at the origin. The de Sitter core is the sign of an antigravity effect. Indeed the matter which generates the spacetime geometry is no longer concentrated at the origin, but it is smeared out with a Gaussian profile. The noncommutative fluctuations sustain this Gaussian profile, preventing its collapse into a Dirac delta. As a result, we cannot speak of a vacuum solution, since Einstein equations have a nonvanishing energy-momentum tensor. Furthermore, this energy-momentum tensor, describing an anisotropic fluid, has a form not equivalent to that of the electromagnetic field stress tensor. For this reason, the application of the Newman-Janis procedure \cite{Drake:1998gf} is no longer straightforward, but requires additional work.

\subsection{New prescription for the complexification}

In this section we extend the Newman-Janis procedure to the case of the line element in (\ref{ncschw}).
The key point is now how to perform the complexification. We need to adapt the general procedure, in order to include the presence of anisotropic stress tensors. Without loss of generality, we can write the line element (\ref{ncschw}) as 
\be
ds^2=\left(1-\frac{2m(r)}{r}\right)du^2+2dudr-d\Omega^2,
\ee
which lets us identify
\be
e^{2 \Phi(r)} = \left(1-\frac{2m(r)}{r}\right) \,\,\,\,  \& \,\,\,\,  \lambda(r) = - \Phi(r).
\ee
Suppose we start from the case $\theta=0$. The line element (\ref{ncschw}) coincides with the Schwarzschild one, $m(r)=M$, and there is no difficulty in following the procedure in \cite{Drake:1998gf}. The mass term in unaffected by the complexification $r\mapsto r^\prime=r+ia\cos\vartheta$ and only the term
\begin{equation}
\frac{1}{r}\mapsto\frac{1}{2}\left(\frac{1}{r^\prime} + \frac{1}{\bar{r^\prime}} \right)=\frac{r}{r^2+a^2\cos^2 \vartheta}
\end{equation}
is modified by the algorithm. As a consequence we write the complexified  metric as
\be
&& e^{2 \Phi(r))}\,\, \mapsto \,\, 1- \frac{2  m({\rm Re}(r^\prime))}{2} \left(\frac{1}{r^\prime} + \frac{1}{\bar{r^\prime}} \right) \nonumber \\
&& \hspace{1.45cm} = 1- \frac{2 m(r) r }{r^2+a^2\cos^2 \vartheta} ,
\label{complmetric}
\ee
where we have expressed $M$ in terms of $m(r^\prime)$ as $M= m({\rm Re}(r^\prime))=m(r)$.  We show now that (\ref{complmetric}) has general validity and does not depend on the specific value $\theta=0$, chosen for the noncommutative parameter. To this purpose, we need to invoke the Kerr-Schild decomposition. As shown in \cite{kerrr}, the Schwarzschild like class of metrics can be cast in the form
\begin{equation}
ds^2=ds_M^2-\frac{f(r)}{r^2}(k_\mu dx^\mu)^2,
\label{kerrschild}
\end{equation}
where $ds_M^2$ is the Minkoswki line element and $k_\mu$ is a lightlike vector in Minkowski coordinates. The function $f(r)$ depends only on the radial coordinate and reads
\begin{equation}
f(r)=2m(r)r.
\label{effe}
\end{equation}
For the conventional Schwarzschild solution we simply have $f(r)=2Mr$. This kind of decomposition permits us to express spinning solutions too. Indeed even if the symmetry has changed, from a spherically symmetric geometry to an axisymmetric geometry, the formal structure of the solution (\ref{kerrschild}) holds
\begin{equation}
ds^2=ds_M^2-\frac{f(r)}{r^\prime\bar r^\prime}(k_\mu dx^\mu)^2,
\label{kerrschild2}
\end{equation}
where $f(r)$ is formally the same as in (\ref{effe}) and $k_\mu$ is expressed in spheroidal coordinates.
As a result we conclude that in all generality the mass term $m(r)$ is unaffected by the complexification and (\ref{complmetric}) is valid for any value of the parameter $\theta$. 

\subsection{The nonsingular line element}
With the above prescription we can follow the Newman-Janis algorithm and write the metric in Boyer-Lindquist coordinates
%namely
%\be
%\Phi(r, \vartheta)=\frac{1}{2}\ln\left(1-\frac{2Mr}{r^2+a^2\cos^2\vartheta}\frac{\gamma(3/2; \frac{r^2+a^2\cos^2\vartheta}{4\theta})}{\Gamma(3/2)}\right)\nonumber \\
%\ee
%The metric in Boyer-Lindquist coordinates is 
%% \begin{eqnarray}
%% g_{\mu\nu}^{(B-L)} =  \left(
%% \begin{array}{cccc}
%%  -\frac{\Sigma (r,\theta )}{a^2 \sin ^2(\theta )+G(r,\theta ) \Sigma (r,\theta )} & 0 & 0 & 0 \\
%%  0 & -\Sigma (r,\theta ) & 0 & 0 \\
%%  0 & 0 & a^2 (G(r,\theta )-2) \sin ^4(\theta )-\sin ^2(\theta ) \Sigma (r,\theta ) & -a (G(r,\theta )-1) \sin ^2(\theta ) %% \\
%%  0 & 0 & -a (G(r,\theta )-1) \sin ^2(\theta ) & G(r,\theta )
%% \end{array}
%% \right)\end{eqnarray}
%% 
%%
\be
&& ds^2_{B-L} = G(r , \vartheta)  dt^2  
- \frac{\Sigma (r,\vartheta )  \, dr^2 }{a^2 \sin ^2 \vartheta +G(r,\theta ) \Sigma (r,\vartheta ) } \nonumber \\
&&       +2  (1 - G(r,\vartheta )) \sin ^2 \theta  \, d t d \phi        -\Sigma (r,\vartheta ) d \vartheta^2
 \nonumber \\
&& 
- \sin ^2 \vartheta  \left[ a^2 (2 - G(r,\vartheta )) \sin ^2 \vartheta  +  \Sigma (r, \vartheta )\right] d \phi^2,
\ee
%
%%
%%
%%
%\begin{widetext}
%\begin{centering}  
%\begin{eqnarray}
%&& g_{\mu\nu}^{(B-L)} =  \left(
%\begin{array}{cccc}
%G(r,\vartheta ) & 0 & 0 & -a (G(r,\vartheta )-1) \sin ^2(\vartheta ) \\
% 0 & -\frac{\Sigma (r,\vartheta )}{a^2 \sin ^2(\vartheta )+G(r,\vartheta ) \Sigma (r,\vartheta )} & 0 & 0 \\
% 0 & 0 &  -\Sigma (r,\vartheta ) & 0 \\
% -a (G(r,\vartheta )-1) \sin ^2(\vartheta ) & 0 & 0 & a^2 (G(r,\vartheta )-2) \sin ^4(\vartheta )-\sin ^2(\vartheta ) \Sigma (r,\vartheta )
%\end{array}
%\right)\nonumber \end{eqnarray}
where 
\begin{eqnarray}
&& \hspace{-0.5cm} G(r, \vartheta) := e^{2 \Phi(r, \vartheta)} = 1-\frac{2Mr}{r^2+a^2\cos^2\vartheta}\frac{\gamma(3/2; r^2/4\theta)}{\Gamma(3/2)} ,  \nonumber \\
&& \hspace{-0.5cm} \Sigma(r, \vartheta) = r^2 + a^2 \cos ^2\vartheta.
%&& a : = a \, \frac{\gamma(3/2; \frac{r^2}{4\theta})}{\Gamma(3/2)}
\end{eqnarray}
%\end{centering}
%\end{widetext}
%\begin{eqnarray}
%R= -\frac{4 \left(\left(2 m^{(1,0)}(r,\theta )+r m^{(2,0)}(r,\theta )\right) \left(-2 r m(r,\theta )+a^2+r^2\right)^2-r^2 m^{(0,1)}(r,\theta
%   )^2\right)}{\left(a^2 \cos (2 \theta )+a^2+2 r^2\right) \left(-2 r m(r,\theta )+a^2+r^2\right)^2}
%\end{eqnarray}
We introduce also the quantity 
\be
\Delta := r^2 - 2 m(r) \,r + a^2
\ee
which is useful to express the metric in a more familiar form
\be
&& ds^2 =  \frac{\Delta - a^2 \sin^2 \vartheta}{\Sigma} dt^2 
 - \frac{ \Sigma}{\Delta }  \, dr^2 
- \Sigma \, d \vartheta^2   \\
&& \hspace{-0.4cm} - \frac{ \Sigma}{\Delta }  \, dr^2 
- \Sigma \, d \vartheta^2 
+  2 a \sin^2 \vartheta \left(1 - \frac{\Delta - a^2 \sin^2 \vartheta}{\Sigma} \right) dt \, d \phi
\nonumber \\
&& \hspace{-0.4cm} -  \, \sin ^2 \vartheta  \left[ \Sigma + a^2 \sin^2 \vartheta \left(2 - \frac{\Delta - a^2 \sin^2\vartheta}{\Sigma}\right)   \right]    d \phi^2 
. \nonumber 
%
%
%\frac{(\Delta - a^2 \sin^2 \theta)}{\Sigma} \left( dt - a    \sin ^2  \theta \, d \phi  \right)^2 + 2 \, a \,  \sin ^2 \theta  
%\, d \phi \left( dt - a 
  % \sin ^2 \theta \, d \phi \right). 
  \label{NCK}
\ee
The above metric coincides with that found in \cite{kerrr}. This is a sign of robustness of our procedure.
In Kerr coordinates the metric reads
\be
&& ds^2_{(K)} = G(r, \vartheta)  du^2 - \Sigma(r , \vartheta) d \vartheta^2 - 2 a \sin^2 \vartheta dr d \phi  \nonumber \\
&& + \left[ a^2 (G(r,\vartheta )-2) \sin ^2 \vartheta - \Sigma (r,\vartheta ) \right] \sin ^2(\vartheta ) d \phi^2 \nonumber \\
&& + 2 a (1-G(r,\theta )) \sin ^2 \vartheta  d \phi \, d u .
\label{core2}
\ee
%
%
%\be
%g_{\mu \nu}^{(K)} = \left(
%\begin{array}{cccc}
% G(r,\vartheta )& 1 & 0 & 0 \\
 %1 & 0 & 0 & -a \sin ^2(\vartheta ) \\
 %0 & 0 & -\Sigma (r,\vartheta ) & a (1-G(r,\vartheta ))
 %  \sin ^2(\vartheta ) \\
% 0 & -a \sin ^2(\vartheta ) & a (1-G(r,\vartheta )) \sin ^2(\vartheta ) & a^2 (G(r,\vartheta )-2) \sin ^4(\vartheta )-\sin ^2(\vartheta ) \Sigma (r,\vartheta )
%\end{array}
%\right)
%\label{core}
%\ee
%
%%
%%
%% \be
%% g_{\mu \nu}^{(K)} = \left(
%% \begin{array}{cccc}
%%  0 & 0 & -a \sin ^2(\vartheta ) & 1 \\
%%  0 & -\Sigma (r,\vartheta ) & 0 & 0 \\
%%  -a \sin ^2(\vartheta ) & 0 & a^2 (G(r,\vartheta )-2) \sin ^4(\vartheta )-\sin ^2(\vartheta ) \Sigma (r,\vartheta ) & a (1-G(r,\vartheta ))
%%    \sin ^2(\vartheta ) \\
%%  1 & 0 & a (1-G(r,\vartheta )) \sin ^2(\vartheta ) & G(r,\vartheta )
%% \end{array}
%% \right)
%% \label{core}
%% \ee
%% 
Here we will not repeat all the discussion about this NCKerr black hole. We just provide some additional clues with respect to the known literature. For instance the curvature singularity is claimed to be removed. %To this purpose the most striking evidence is the value of the Kretschmann invariant at the origin [see ()].
%\subsection{Singularity}
To attack the singularity problem we can study not only the Ricci scalar but also the Kretschmann invariant 
defined as $K(r, \vartheta) = R_{\mu \nu \rho \sigma} R^{\mu \nu \rho \sigma}$.
The Ricci scalar, 
\be
R(r , \vartheta) = \frac{M r^2 e^{-\frac{r^2}{4 \theta}} \left(r^2-8 \theta\right)}{\sqrt{\pi } \, \theta^{5/2} \,   \left(a^2 \cos (2 \vartheta )+a^2+2 r^2\right)},
\label{ricciscalar}
\ee
is regular in $r=0$, $\vartheta = \pi/2$ but it assumes two different values depending of the way one reaches the origin. This is the signature of the presence of the so called ``de Sitter belt''. If we approach the origin moving on the equatorial plane, i.e. $\vartheta=\pi/2$, we find a regular rotating de Sitter geometry
\be
&&  \lim_{r  \rightarrow 0} \,  \left( \lim_{\vartheta \rightarrow \pi/2} R(r, \vartheta) \right) = - \frac{4 M }{\sqrt{ \pi} \, \theta^{3/2}}.
\ee
However if we approach the origin along an arbitrary plane, i.e. $r\to0$ with $\vartheta\neq\pi/2$, we reach the equatorial disk. Since in this region there is no matter, we find a Minkowski geometry, namely
\be
&& \lim_{\vartheta  \rightarrow \pi/2} \, \left( \lim_{r \rightarrow 0} R(r, \vartheta) \right) = 0.  %\\
%&&  \lim_{r  \rightarrow 0} \,  \left( \lim_{\vartheta \rightarrow \pi/2} R(r, \vartheta) \right) = - \frac{4 M }{\sqrt{ \pi} \, \ell^3}.
\ee
This feature can also be seen by studying the Kretschmann invariant, whose general formula is given in (\ref{KI}). Again we obtain two different values according to the way we approach the origin
\begin{eqnarray}
&& \lim_{r \rightarrow 0} \, \left( \lim_{\vartheta \rightarrow \pi/2} K(r,\vartheta) \right) = \frac{8 M^2}{3 \pi \theta^3} , 
 \\
&& \lim_{\vartheta \rightarrow \pi/2} \, \left(\lim_{r \rightarrow 0} K(r,\vartheta) \right)= 0. \nn
\end{eqnarray}
%% Expanding $K(r, \vartheta)$ for $\vartheta \rightarrow \pi/2$ and small $r$ we find
%% \be
%% \lim_{\vartheta \rightarrow \pi/2} K(r , \vartheta) = 
%% \frac{8 M^2}{3 \pi  \ell^6}-\frac{2 M^2 r^2}{\pi  \ell^8}+\frac{259 M^2 r^4}{300 \pi  \ell^{10}}-\frac{121 M^2 r^6}{504 \pi \ell^{12} }+\frac{3299 M^2 r^8}{70560 \pi  \ell^{14}}-\frac{65 M^2 r^{10}}{9504 \pi 
%%    \ell^{16} }+O\left(r^{11}\right)
%% \ee
To clarify this point it is worthwhile to expand the metric for small $r$, keeping $\vartheta = \pi/2$ 
\be
&& ds^2 %= \left( 1 - \frac{2 m(r)}{r} \right) dt^2 - \frac{dr^2}{ 1 -  \frac{ 2 m(r) r -a^2}{r^2} } +  \frac{2 a m(r) }{r} dt d \phi - \left( r^2 + a^2 + \frac{2 a^2 m(r)}{r}  \right) d \phi^2 \nn \\ 
%&& \hspace{0.6cm}
\approx \left( 1 - \frac{\Lambda r^2}{3} \right)\  dt^2 - \frac{r^2}{ a^2+ r^2 %- \frac{ \Lambda r^4}{3} 
}\  dr^2 \nonumber \\
&& + 2 \frac{\Lambda a r^2}{3}\  dt \ d \phi 
- \left(   r^2 +a^2  + \frac{ \Lambda a^2 r^2}{3} \right) \ d \phi^2 , 
\ee
which corresponds to a rotating de Sitter geometry with $\Lambda=M/\theta^{3/2}\sqrt\pi$. If one approaches the origin for $\vartheta\neq\pi/2$, one finds the flat disk
%the last approximation is for $r \rightarrow 0$.
\begin{equation}
ds^2\approx dt^2 - \cos^2\vartheta dr^2-a^2\sin^2\vartheta d \phi^2.
\end{equation}
The two geometries have been explained in terms of a rotating string, which replaces the ring singularity \cite{kerrr}. This is equivalent to saying that instead of the conventional infinite discontinuity in the curvature, we have a finite jump, i.e. $4\Lambda$, thanks to the presence of the de Sitter belt. 

\subsection{Thermodynamics}

\begin{figure}
  \begin{center}
    \includegraphics[height=6.5cm]{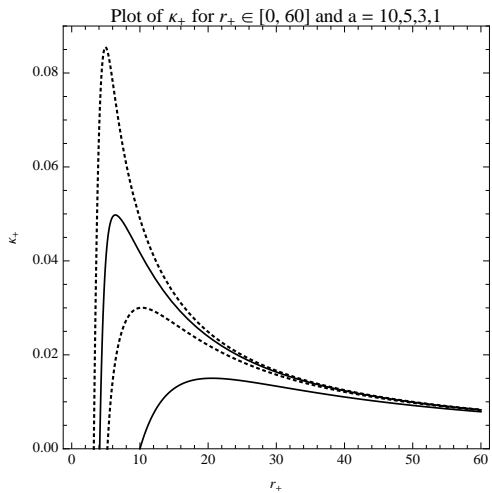}
   \end{center}
   \caption{\label{Kerrtemp} 
   Plot of the surface gravity as a function of the horizon radius for different values of the $a$. The rotation has just the effect of increasing the values of the mass of the black hole remnant.
   } 
   \end{figure}

To complete the analysis of the NCKerr solution, we face the problem of the Hawking temperature, a fact yet unexplored in the literature. To define the Hawking temperature we need the surface gravity $\kappa$ in order to have
\begin{eqnarray}
T=\frac{\hbar c
\kappa}{2\pi k_{B}},  
\end{eqnarray}
where we have temporarily restored all the constants for sake of clarity. The surface gravity is defined by
\be
\kappa^2 = - \frac{1}{2} \nabla^{\mu} \chi^{\nu} \nabla_{\mu} \chi_{\nu} , 
\ee
where $ \chi^{\nu}$ are null Killing vectors. In the case of an axisymmetric geometry, the quest of null Killing vectors requires a certain attention. We start from 
\be 
\chi^{\mu} = \xi^{\mu} + \Omega \, m^{\mu} , 
\ee
where $\xi^{\mu} \equiv\partial_{t}$ and $m^{\mu}\equiv \partial_{\phi}$ are, respectively, 
the Killing vector associated with the time translation invariance and rotational invariance.
We find $\Omega$ imposing $\chi^{\mu}$ to be a null vector, i.e.
\be 
&&
\chi^{\mu} \chi_{\mu} = g_{tt} + 2 \Omega g_{t \phi} + \Omega^2 g_{\phi \phi} =0 %\,\,\,\,  \Longrightarrow 
 \\
&& \hspace{-0.4cm} \Longrightarrow 
\Omega = - \frac{ g_{t \phi}}{g_{\phi \phi}} \pm \sqrt{ \left(\frac{ g_{t \phi}}{g_{\phi \phi}} \right)^2 - 
\frac{ g_{t t}}{g_{\phi \phi}} } = \omega \pm \sqrt{ \omega^2 - 
\frac{ g_{t t}}{g_{\phi \phi}} },  \nonumber 
\label{nullcond}
\ee
where 
\be
 \hspace{-0.5cm} \omega =  \frac{ g_{t \phi}}{g_{\phi \phi}} = 
 \frac{ - a \left[ \Sigma - (\Delta - a^2 \sin^2 \vartheta) 
\right]}{\Sigma^2 + a^2 \sin^2 \vartheta \left[ 2 \Sigma - (\Delta - a^2 \sin^2 \vartheta) \right]}, 
\ee
and therefore
\be
\Omega = \omega \pm \frac{\Sigma \, \Delta^{\frac{1}{2}}}{\sin \vartheta \left[ ( a^2 \sin^2 \vartheta+ \Sigma)^2 - \Delta a^2 \sin^2 \vartheta \right]} .
\ee
On the event horizon $\Delta = 0$. Thus $\Omega$ reduces to 
$\Omega_H = \omega(r_+)= a/(r_+^2 + a^2)$,
where $r_+$ is  implicitly defined by $\Delta(r_+) = 0$.
The surface gravity for a general function $\Delta(r)$ is 
\be
\kappa = \frac{\Delta'(r_+)}{2 (r_+^2 + a^2)} 
\label{kappa}
\ee
where $'$ denotes the derivative respect to $r$.
More specifically when $\Delta(r)=r^2+a^2-2m(r)r$ as in the NCKerr spacetime  we find 
\be 
\kappa = \frac{r_+}{2 (r_+^2 + a^2)} \left[ 1 - \frac{a^2}{r_+^2} - \frac{( r_+^2 + a^2) r_+}{ 4 \theta^{3/2}} 
\frac{ e^{ - r_+^2/4 \theta}}{ \gamma(3/2; r_+^2/4 \theta)} \right].
\label{kexply} \nonumber 
\ee
As a result the temperature becomes a function $T=T(r_+, a)$. We notice that for having a positive defined temperature $a$ cannot exceed $r_+$. In Fig. \ref{Kerrtemp}, we see that the profile of the temperature is roughly equivalent to that of the nonrotating solution. After a temperature maximum, the black hole cools down to a zero temperature black hole remnant final state. The size and the mass of this remnant have increased with respect to the nonrotating case, since also the rotational kinetic energy is stored in the final configuration. However, this scenario does not take into account the loss of angular momentum due to the Hawking emission and the consequent transition into a Schwarzschild phase. In other words the black hole evaporation is a multi phase process that would require further investigations.

%%%%%%%%
\section{Regular Kerr-Newman black hole}
\label{kerrrn}

\subsection{The nonsingular line element}
After the foreplay about the NCKerr solution, we can make a step forward. We want to derive a new spacetime geometry which comprises all the three fundamental black hole parameters, mass, charge and angular momentum. In the spirit of what we found in the previous section, we would like the new solution to be singularity free for the presence of an effective minimal length. This would close the program of the NCBHs, initiated in 2005 with the NCSchw solution.  To tackle this problem we recall the noncommutative geometry inspired Reissner-Nordstr\"{o}m (NCRN) solution described by the line element \cite{Ansoldi:2006vg}
\be
ds^2=e^{2 \Phi(r)} dt^2 -\frac{dr^2}{e^{2 \Phi(r)}}-r^2 d\Omega^2 , 
\ee
where $\exp(2 \Phi)$ is the following more involved but regular quantity 
\begin{widetext}
\begin{centering} 
\be
%&& 
\hspace{-0.5cm}e^{2 \Phi(r)} %:= 1 - \frac{2 m(r)}{r} + \frac{q(r)^2}{r^2} \nonumber \\
%&&
 = 1-\frac{2}{r} \,   \underbrace{\left[\frac{ M \gamma(3/2;\ r^2/4\theta)}{\Gamma(3/2)} \right]}_{m(r)} +\frac{1}{r^2} 
\underbrace{\left\{\frac{Q^2}{\pi}
\left[\gamma^2 (1/2; r^2/4\theta)-\frac{r}{\sqrt{2\theta}}\gamma(1/2; r^2/2\theta) + r\ \sqrt{\frac{2}{\theta}}\gamma(3/2;\ r^2/4\theta)\right] \right\}}_{q(r)^2}.
\ee
\end{centering}
\end{widetext}
This geometry is generated by an energy-momentum tensor made up of two parts
\begin{equation}
T^{\mu\nu}=T^{\mu\nu}_{matt.} + T^{\mu\nu}_{el.} \, , 
\end{equation}
where $T^{\mu\nu}_{matt.}$ describes an anisotropic neutral fluid and $T^{\mu\nu}_{el.}$ is formally the usual electromagnetic field stress tensor. However, the Maxwell field $F^{\mu\nu}$ solves the equation
\begin{equation}
\frac{1}{\sqrt{-g}}\, \partial_\mu\,\left(\, \sqrt{-g}\, F^{\mu\nu}\, \right)= J^\nu \, , 
\end{equation}
whose source term is subjected to the noncommutative smearing effect, i.e. the presence of a minimal length. As a consequence the resulting electric field is
\begin{equation}
E\left( r \right)=\frac{2{\cal Q}}{\sqrt\pi\, r^2}\, \gamma\left(\, \frac{3}{2}\ ; \frac{r^2}{4\theta}\,\right) \,  ,
\label{clmb}
\end{equation}
where ${\cal Q}=Q\ c^2/G^{1/2}$ is the electric charge. Another key feature of the solution is its ADM mass $M$, which includes also the regularized electrostatic self-energy of the system
\begin{equation}
M =\oint_\Sigma d\sigma^\mu\left(\, T_\mu^0\vert_{matt.} + T_\mu^0\vert_{el.}  \,\right) \label{mtot} \,  ,
\end{equation}
where $\Sigma$ is a $t=\mathrm{const.}$, closed three-surface at infinity. In the conventional Reisnerr-Nordst\"{o}m solution this term diverges and it is simply disregarded. Indeed to avoid the problem of the singularity, one solves Einstein equations in a restricted domain $\mathbb{R}\times\left(\mathbb{R}^3  \setminus \left\{0\right\}\right)$, i.e. in vacuum, and identifies the two integration constants with mass and charge from the behavior of the solution at infinity. 

\begin{figure}
  \begin{center}
    \includegraphics[height=6.5cm]{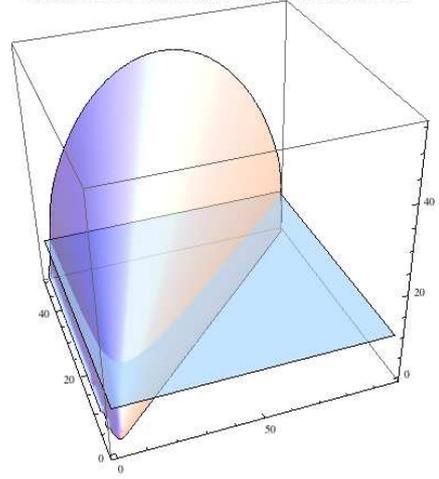}
   \end{center}
   \caption{\label{horizonplot} Plot of the function $\Delta=0$ for the neutral rotating case. The parameter $a$ is along  the z-axis there, the horizon radius is along the x-axis and the mass $M$ is along the y-axis. Qualitatively this is the behaviors of the rotating charges solution too.} 
 %\label{Ricci}
   \end{figure}

\begin{figure}
  \begin{center}
   \includegraphics[height=5cm]{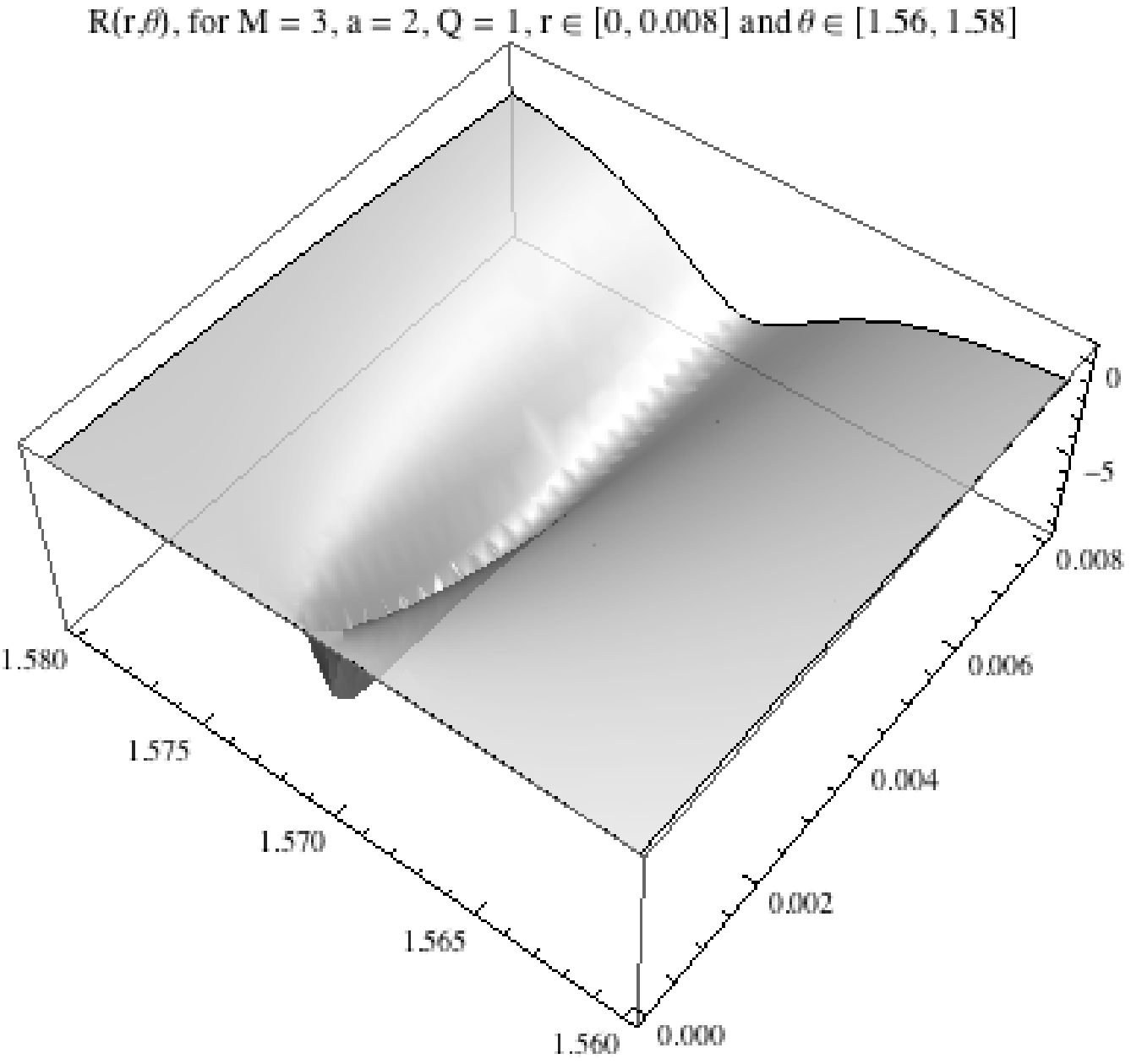}
   \includegraphics[height=5cm]{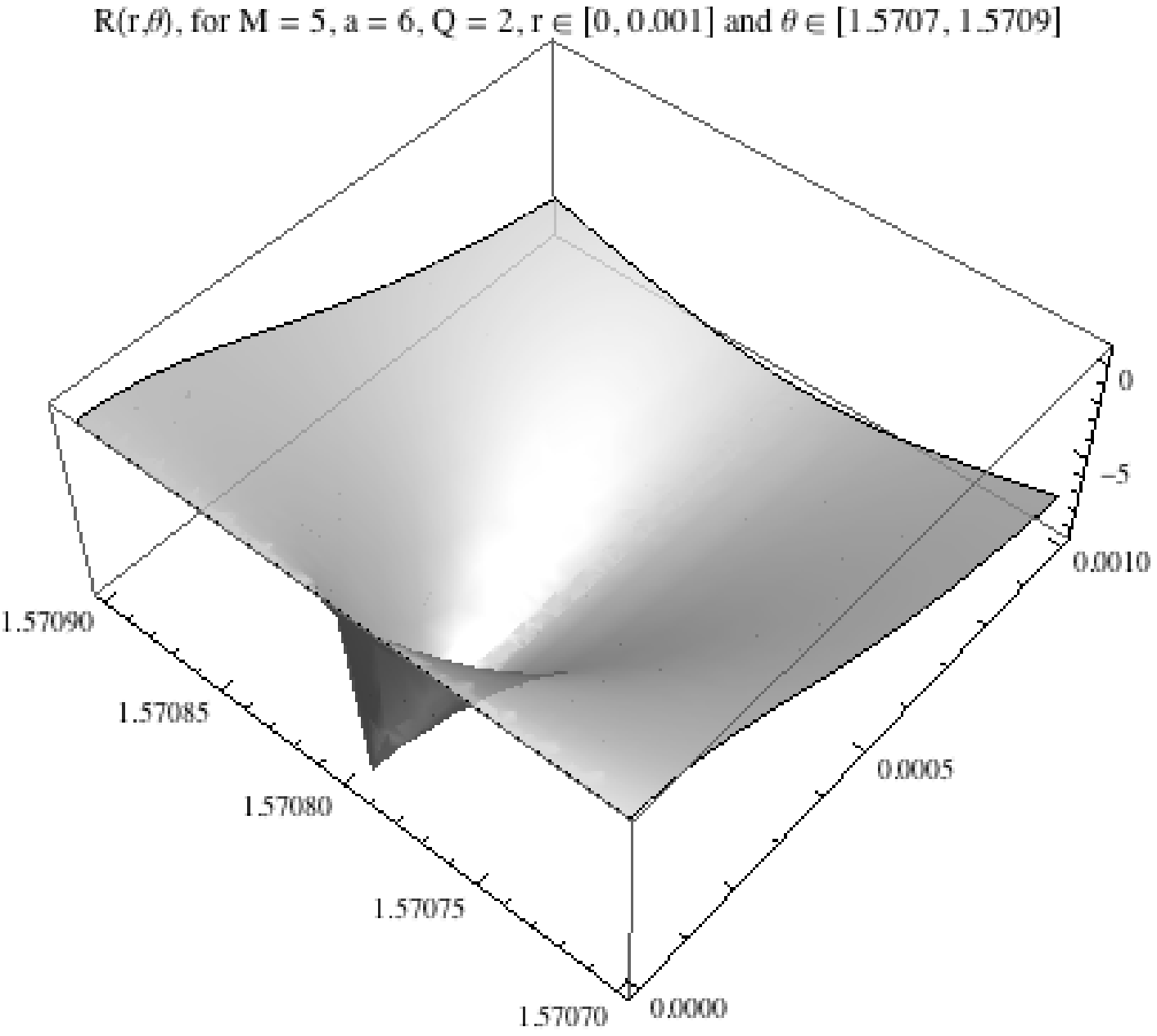}
    \includegraphics[height=4.5cm]{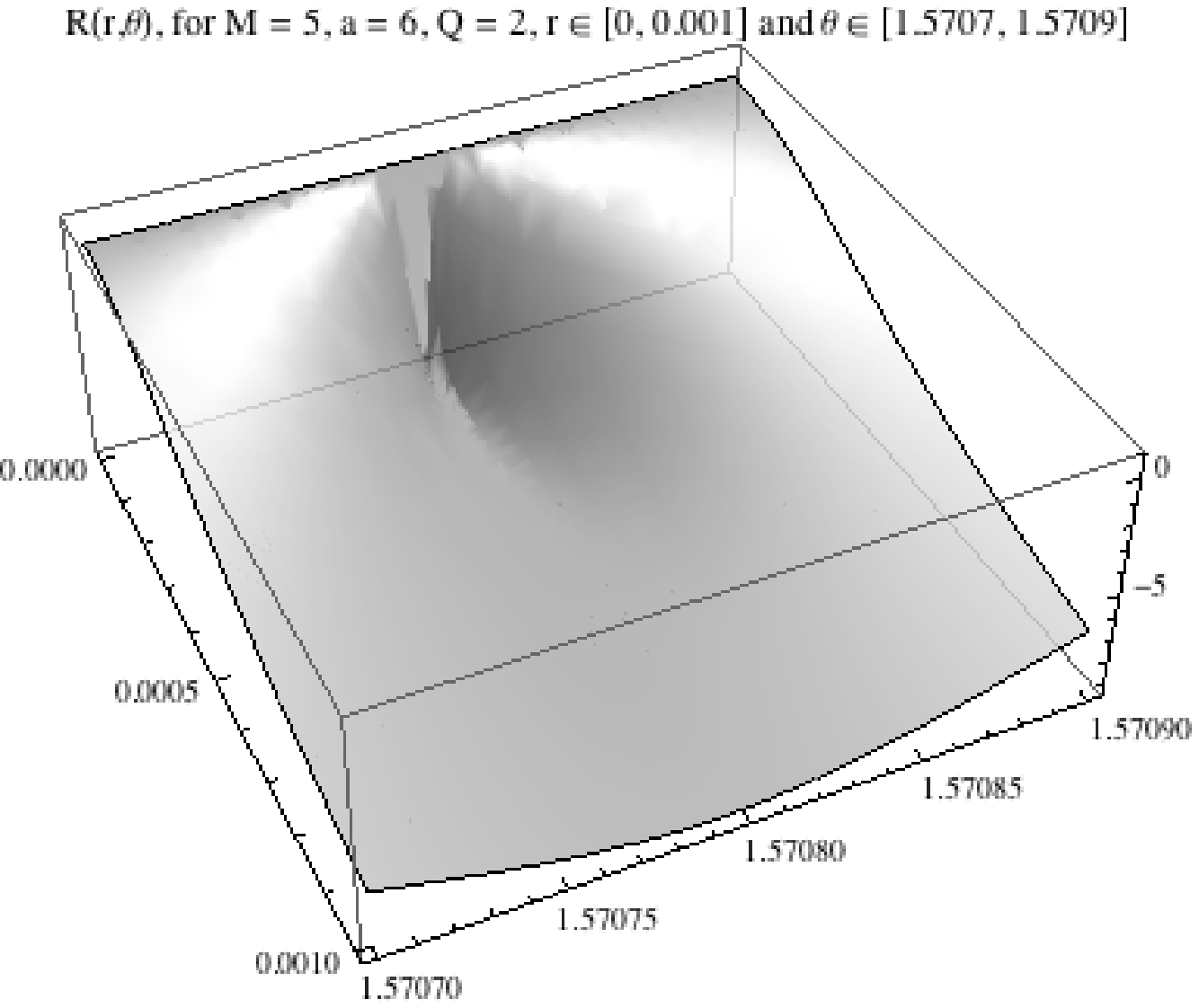}
   \end{center}
   \caption{\label{RicciScalarPlot} Plot of the Ricci scalar as a function or $r$ and $\vartheta$ for different values of the couple $(M, a)$. Along the equatorial plane, i.e. $\vartheta=\pi/2\sim 1.57$ the de Sitter belt appears. For all other angles, one finds a flat disk i.e. vanishing curvature.
   } 
 \label{Ricci}
   \end{figure}

Given the form of the electric field (\ref{clmb}), the actual form of $T_\mu^0\vert_{el.}$ turns out to be rather different from the usual electromagnetic stress tensor. Therefore the Newman-Janis algorithm must be reviewed. As in the previous sections we invoke the validity of the Kerr-Schild decomposition. The conventional Reissner-Nordstr\"{o}m solution can be written as
\begin{equation}
ds^2=ds_M^2-\frac{f(r)}{r^2}(k_\mu dx^\mu)^2,
\label{kerrschild3}
\end{equation}
where the function $f(r)=2Mr-Q^2$. The NC geometry inspired solution has the same form. The delocalization of the source term in Einstein equations has just the effect of modifying the function $f(r)$, which now is $f(r)=2m(r)r-q(r)^2$. This is in the spirit of what we have seen for the Schwarzschild case. The function $f(r)$ is mapped into $f(r)=2m(r)r$, passing from the conventional to the NC geometry inspired case, but maintains the structure $f\propto r\times \ \mathrm {mass}$. In the charged case the only difference is that the ``mass'' contains also the energy stored in the field, namely
\begin{equation}
2\ \mathrm {mass}\mapsto 2\ \mathrm {mass}-\frac{\mathrm{charge}^2}{r}.
\label{massch}
\end{equation}
As a result for the class of all rotating geometries we still have (\ref{kerrschild2}). We conclude that the function $f(r)$ is formally unaffected even if expressed in spheroidal coordinates rather than spherical ones.
Therefore we can extend the validity of our prescription about the complexification of line elements not comprised by the Newman-Janis algorithm in order to include the NCRN geometry.
%where 
%\be
%F(r) = \gamma^2 (1/2; r^2/4\theta)-\frac{r}{\sqrt{2\theta}}\gamma(1/2; r^2/2\theta)
%\ee
For this reason we write
\be
&& e^{2 \Phi(r)} = 1 - \frac{2 m(r)}{r} + \frac{q(r)^2}{r^2} \nonumber \\
&& 
\mapsto \, 
1 - \left( \frac{1}{r'}+ \frac{1}{\bar{r'}} \right) \, m[ {\rm Re} (r') ]  + \frac{1}{r' \bar{r'}} \, q^2[ {\rm Re} (r') ]  \nonumber \\
&& \hspace{0.15cm}= 1 - \frac{2 m(r) r}{r^2+a^2\cos^2(\vartheta)} + \frac{q(r)^2}{r^2+a^2\cos^2(\vartheta)}.
\ee
%%\be
%%\lambda(r)=-\Phi(r)
%%\ee
The line element is formally equivalent to the neutral one
\be
&& ds^2 =  \frac{\Delta - a^2 \sin^2 \vartheta}{\Sigma} dt^2 
- \frac{ \Sigma}{\Delta }  \, dr^2 
- \Sigma \, d \vartheta^2
\\
&&
 +  2 a \sin^2 \vartheta \left(1 - \frac{\Delta - a^2 \sin^2 \vartheta}{\Sigma} \right) dt \, d \phi \nonumber \\
&&  
 -  \, \sin ^2 \vartheta  \left[ \Sigma + a^2 \sin^2 \vartheta \left(2 - \frac{\Delta - a^2 \sin^2\vartheta}{\Sigma}\right)   \right]    d \phi^2. \nonumber \\
%
%
%\frac{(\Delta - a^2 \sin^2 \theta)}{\Sigma} \left( dt - a    \sin ^2  \theta \, d \phi  \right)^2 + 2 \, a \,  \sin ^2 \theta  
%\, d \phi \left( dt - a 
  % \sin ^2 \theta \, d \phi \right). 
  \label{NCKN}
  \ee
apart from the function $\Delta(r)$ 
\be
\Delta = r^2 - 2 m(r) r + q(r)^2 + a^2,
\ee
which now depends on the ``charged'' $f(r)$. Because of the equivalent formal structure of the line element, considerations about the horizons are the same as in the neutral case. The horizon equation $1/g_{rr}=0$, gives $\Delta(r_H)=0$. Again this equation cannot be solved in a closed form as $r_H=r_H(M, Q, a)$. Anyway the horizon equation is formally equivalent to its corresponding one for the nonrotating case. There is just an extra parameter $a^2$ which simply shifts the roots of the equation. As a result we have the following scenario (see Fig. \ref{horizonplot})
\begin{enumerate}
\item for $M>M_{extr.}$ there are two distinct horizons $r_\pm$, corresponding to the case of nonextremal black hole;
\item for $M=M_{extr.}$ there is one degenerate horizon $r_{extr.}$, corresponding to the case of extremal black hole;
\item for $M<M_{extr.}$ there is no horizon and the line element describes the regular geometry of a charged spinning object. Extending the terminology already in use we will speak of charged spinning mini-gravastar, a system completely governed by the quantum mechanical fluctuations of the manifold.
\end{enumerate} 
In support of our claim about the regularity of the spacetime, we must study the singularity problem. %studying the Rocci scalar. %which classically 
%and diverges in $r \rightarrow 0$, $\theta \rightarrow 0$ because the solution is not a vacuum 
%solution.
The Ricci scalar as function of $r$ and $\vartheta$ is 
\be
R(r, \vartheta) = \frac{-4 r m''(r)-8 m'(r)+2 [q^{2}]''(r)}{a^2 \cos (2 \vartheta )+a^2+2 r^2} \, , 
\label{ricciKN}
\ee
introducing $m(r)$ and $q(r)^2$ we find 
\be
&& \hspace{-0.5cm} R(r, \vartheta) = \frac{e^{-\frac{r^2}{2 \theta}}}{4 \pi 
   \theta^3 \ r \left(a^2 \cos (2 \vartheta )+a^2+2 r^2\right)}  \times \\
  && \hspace{-0.5cm} 
   \Big[ 4 \sqrt{\pi } \sqrt{\theta} M r^5
   e^{\frac{r^2}{4 \theta}} 
   -\sqrt{2} Q^2 r^5 e^{\frac{r^2}{4 \theta}}+8 \theta Q^2 r^3 \left(\sqrt{2}
   e^{\frac{r^2}{4 \theta}}+1\right) \nn \\
  && \hspace{-0.5cm} 
   -8 \sqrt{\pi } \theta^{\frac{3}{2}} r^2 e^{\frac{r^2}{4 \theta}} \left(4 M r+Q^2\right) \nn 
   +8
   \theta^{\frac{3}{2}} Q^2 r^2 e^{\frac{r^2}{4 \theta}} \, \Gamma \left( \frac{1}{2}, \frac{r^2}{4 \theta}   
 \right) \Big]. \nonumber
\ee
The plot for the Ricci scalar, with the analytical limits near the classical singularity, is in Fig.\ref{RicciScalarPlot}.
As in the neutral case, a jump of the Ricci scalar at the origin appears 
\be
&& \lim_{r \rightarrow 0} \lim_{\vartheta \rightarrow \pi/2} R(r, \vartheta) = \frac{\sqrt{2} Q^2-4 \sqrt{\pi } \sqrt{\theta} M}{\pi  \theta^2}, \label{chbelt} \nn \\
&& \hspace{-0.1cm} \lim_{\vartheta \rightarrow \pi/2} \lim_{r \rightarrow 0} R(r, \vartheta) = 0.
\ee
These limits deserve further consideration with respect to the new case. Indeed we have seen that for the NCKerr solution approaching the origin on the equatorial plane lets us experience a cosmological term  $\Lambda=M/\theta^{3/2}\sqrt\pi$. This is exactly the value seen when approaching the origin of the corresponding spherically symmetric solution, namely, the NCSchw black hole. The only difference in the spinning case is that such a $\Lambda$ emerges only for limits along the equatorial plane. But there is something more. The same value emerges as far as one approaches the origin of the NCRN solution. More precisely in the latter case only the bare mass term 
\begin{equation}
M_0=\oint_\Sigma d\sigma^\mu \  T_\mu^0\vert_{matt.} \,  ,
\end{equation}
contributes to the effective cosmological constant, while the electromagnetic contribution turns out to be subleading. This would imply that also in our new spinning charged solution the value of $\Lambda$ could be proportional to $M_0$ as in the nonrotating case. Indeed if we give for granted the scheme in (\ref{massch}), we can exploit the properties of the NCRN solution, for which the mass term $M$ reads
\begin{equation}
M=M_0+\frac{Q^2}{\sqrt2\theta\pi}.
\label{massen}
\end{equation}
We notice that by inserting the above relation in (\ref{chbelt}), the electromagnetic energy exactly cancels out,  confirming our line of reasoning
\begin{equation}
 \lim_{r \rightarrow 0} \lim_{\vartheta \rightarrow \pi/2} R(r, \vartheta) = -\frac{4 M_0}{\sqrt{\pi}  \theta^{3/2}}.
\end{equation}
This feature discloses new insights about the nature of the energy-momentum tensor. 
%% The components of the metric $g_{tt}$ and $g_{\phi \phi}$, more precisely 
%% \be 
%% &&
%%  \lim_{r  \rightarrow 0^+} \left( \lim_{\vartheta \rightarrow \pi/2} g_{tt} \right) = 1 \,\,\, , \hspace{1cm}
%%  %\left(\frac{2 Q^2}{\pi  L^2}-1\right) \nn \\
%% %
%% \lim_{r  \rightarrow 0^-} \left( \lim_{\vartheta \rightarrow \pi/2} g_{t t} \right) = 1 +  \frac{2 Q^2}{\pi  L\ell^2}  \nn %% \\
%% &&
%%  \lim_{r  \rightarrow 0^+} \left( \lim_{\vartheta \rightarrow \pi/2} g_{\phi \phi} \right) = - a^2 
%%  %
%%  \,\,\, , \hspace{1cm}
%% \lim_{r  \rightarrow 0^-} \left( \lim_{\vartheta \rightarrow \pi/2} g_{\phi  \phi} \right) = - a^2 \left(- \frac{2 Q^2}{\pi  %% \ell^2} + 1\right) \nn
%% \ee
%% 
%% %Then one can follow the general procedure in \cite{Drake:1998gf} and gets the corresponding axisymmetric line element.
%% 
\subsection{Stress-energy tensor}
Before addressing the problem of the energy-momentum tensor we recall some properties of the new solution.
The charged spinning geometry has been simply mapped from the nonrotating one, without formally modifying the distributions of mass and charge. Furthermore, the charged spinning geometry can also be mapped from the neutral spinning one, simply redefining $f(r)$ or $\Delta$, in order to include a charge term. We conclude that the Newman-Janis algorithm does not affect the general form of the stress tensor, that will be 
\begin{equation}
T^\mu_{\nu}=( \rho+p_\vartheta )\ ( u^\mu u_\nu-\ell^\mu\ell_\nu )-p_\vartheta\ \delta^\mu_\nu.
\end{equation}
The above expression can be split in two contributions as in the NCRN case
\begin{equation}
T^\mu_{\nu}=T^\mu_{\nu}\vert_{matt.}+T^\mu_{\nu}\vert_{el.} \, , 
\end{equation}
where $T^\mu_{\nu}\vert_{matt.}$ is the source term for the neutral spinning solution, namely
\begin{equation}
T^\mu_{\nu}\vert_{matt.}=( \rho_0+(p_\vartheta)_0 )\ ( u^\mu u_\nu-\ell^\mu\ell_\nu  )-(p_\vartheta)_0\ \delta^\mu_\nu.
\end{equation}
We start from $T_0^0$ which represents the energy density of the system. We have already obtained a prescription for it from the form of the total mass energy of the system (\ref{massen}). If we really believe that the charged solution can be obtained from the neutral one assuming a new form for $f(r)$, the resulting energy density can be obtained invoking the validity of (\ref{massch}). Therefore it reads
\begin{equation}
T_0^0\equiv\rho(r)=\rho_m(r)+\rho_q(r) \, ,
\end{equation}
where $\rho_M(r)$ contains both the term leading to the bare mass $M_0$ and the electrostatic self-energy density
\begin{equation}
\rho_m(r)=\rho_0(r)+\rho_{\mathrm{self}}(r)
\end{equation}
and $\rho_q(r)$ can be implicitly defined through
\begin{equation}
2\pi\int_0^r \ dr\int_0^\pi\ d\vartheta\sin\vartheta\ \frac{\Sigma^2}{r^2}\ \rho_q(x)=-\frac{1}{2}\frac{q(r)^2}{r}.
\end{equation}
Here the integration volume in spheroidal coordinates is considered. 
%%This leads to
%%\begin{equation}
%%\rho_q(r)=\frac{1}{2}\left(\frac{q^2}{r}-\frac{2qq'}{r}\right).
%%\end{equation}
In agreement with \cite{kerrr}, $\rho(r)$ is an invariant energy density %$\rho(r)=T_{\mu\nu}u^\mu u^\nu$ 
and the function $\rho_m(r)$ is connected to the mass distribution as
\begin{equation}
\rho_m(r)=\frac{r^4}{\Sigma^2}\ \rho_G(r),
\end{equation}
where $\rho_G(r)$ is the original spherically symmetric Gaussian in agreement with the basic motivations of all NC geometry inspired solutions
\begin{equation}
\rho_G(r)=\frac{M}{(4\pi\theta)^{3/2}}\ e^{-r^2/4\theta}.
\end{equation}
To check this result we need to prove that (\ref{massen}) is satisfied. To this purpose we recall that the mass $M$
can be obtained from
\begin{equation}
M=2\pi\int_0^\infty\ dr\int_0^\pi\ d\vartheta\sin\vartheta\ \frac{\Sigma^2}{r^2}\ \rho(r). 
\end{equation} 
The integral can be written as
\begin{equation}
M=2\pi\int_0^\infty dr \int_0^\pi d \vartheta\sin\vartheta \frac{\Sigma^2}{r^2} \left[\rho_0(r) + \rho_{\mathrm{self}}(r) +\rho_q(r)\right]. \nonumber
\end{equation}
The integration of $\rho_q(r)$ gives $\sim Q^2/r$ which vanishes at infinity. Therefore the integral becomes
\begin{equation}
M=4\pi\int_0^\infty r^2  \rho_G(r),
\end{equation}
where $\rho_G(r)$ is nothing but the profile of the  NCRN case. Therefore we know that its integration leads to a mass $M$ including the regularized electrostatic self energy as in (\ref{massen}).
The other entries of the stress tensor are connected through Einstein equations to the generic functions $f(r)$ and $\Delta$
\begin{eqnarray}
\frac{d^2f}{dr^2}&=&16\pi g_{\vartheta\vartheta}\left(T_r^r+T_\vartheta^\vartheta\right) , \\
\frac{d^2\Delta}{dr^2}&=&2-16\pi g_{\vartheta\vartheta}\left(T_r^r+T_\vartheta^\vartheta\right).
\end{eqnarray}
We recall that also these functions can be written as $f(r)=f_0(r)+f_{el.}(r)$, $\Delta=\Delta_0+\Delta_{el.}$, with $f_{el.}(r)=-q(r)^2$ and $ \Delta_{el.}=q(r)^2$. As a result we have
\begin{equation}
-\frac{d^2 [q^2]}{dr^2}=-16\pi\left(a^2\frac{\sin^4\vartheta}{\Sigma}\ q^2\right)\left(T_r^r+T_\vartheta^\vartheta\right)\vert_{el.} \, .
\end{equation}
The remaining terms can be computed requiring the conservation of the stress tensor. 
The analytical components are given in Appendix C.
\subsection{Thermodynamics}

The thermodynamics of the NC geometry inspired Kerr-Newman black hole can be studied in analogy to what we found for the neutral spinning case. We recall that the surface gravity for a general function $\Delta(r)$ is 
\be
\kappa = \frac{\Delta'(r_+)}{2 (r_+^2 + a^2)} .
\ee
When $Q \neq 0$ we have that $\Delta=r^2+a^2-2m(r)r+q(r)^2$ and thus 
%\begin{widetext}
\be 
&& \hspace{-0.7cm} \kappa = \frac{r_+}{2 (r_+^2 + a^2)} \Big[ 1 - \frac{a^2}{r_+^2} - \frac{q^2(r_+)}{r_+^2} 
+ \frac{[q^2]' (r_+)}{r_+}\nn \\
&& - \frac{( r_+^2 + a^2 + q^2(r_+)) r_+}{ 4 \theta^{3/2}} 
\frac{ e^{ - r_+^2/4 \theta}}{ \gamma(3/2; r_+^2/4 \theta)} 
  \Big]. 
\label{kexplyQ}
\ee
%\end{widetext}
We can write the above formula as
\be 
&& \hspace{-1cm}
\kappa = \kappa_0+\frac{r_+}{2 (r_+^2 + a^2)} \left[  - \frac{q^2(r_+)}{r_+^2} 
+ \frac{[q^2]' (r_+)}{r_+} \right. \nn\\
&& 
\left.- \frac{(q^2(r_+)) r_+}{ 4 \theta^{3/2}} 
\frac{ e^{ - r_+^2/4 \theta}}{ \gamma(3/2; r_+^2/4 \theta)} 
  \right],
\ee
where $\kappa_0$ is the temperature in the neutral rotating case.
As in the NCRN case, the presence of the charge just lowers down the temperature with respect to the neutral case. 
Therefore we expect a temperature profile that qualitatively resembles what found in Fig. \ref{Kerrtemp} for the neutral case.

 %%\begin{figure}
%%  \begin{center}
 %%   \includegraphics[height=4.5cm]{kappapiu.eps}
 %%  \end{center}
 %%  \caption{%\label{RicciScalarPlot} Plot of the Ricci scalar for different values of the couple $(m, a)$.
 %%  } 
%% \label{Ricci}
  %% \end{figure}

\section{Higher-dimensional rotating black holes} %rotating black holes}
\label{extrakerrr}
\subsection{Higher-dimensional regular Kerr black hole} %rotating black holes}
In view of possible appearance of microscopic black hole in TeV gravity experiments it is mandatory to extend our analysis to the extradimensional scenario. We start from the spherically 
symmetric black hole 
\be 
&& ds^2_{(d+1)} =\underbrace{ \left(1 - \frac{2 \mu(r)}{r^{d-2}} \right)}_{G(r)} dt^2 - \frac{dr ^2}{\left(1 - \frac{2 \mu(r)}{r^{d-2}} \right)}
\nonumber \\
&& - \underbrace{r^2}_{g_{\Omega^{2}}} (d \vartheta^2 + \sin^2 \vartheta \phi^2) \
- \underbrace{ r^2 \cos^2 \vartheta}_{g_{\Omega^{d -3}}} d \Omega^{d-3}, \nn \\
&& \mu(r) = \frac{M}{M_{*}^{d -1} \Gamma(d/2)} \gamma(d/2; r^2/(4 \ell^2).
\ee
 and we apply our usual prescription for the complexification, i.e.
\be 
&& G(r) = \left(1 - \frac{2 \mu(r)}{r^2 r^{d-4}} \right) \mapsto 
\left(1 - \frac{2 \mu[(r'+\bar{r}')/2]}{r' \bar{r}' \,  [(r' + \bar{r}' )/2]^{d-4}}\right) \nonumber \\
&& \hspace{0.8cm} 
=  1-  \frac{ 2 \mu(r)}{R^2 \,  r^{d-4}}, \nn \\
 && g_{\Omega^{2}} \mapsto r' \bar{r}' = R^2, \nn \\
&& g_{\Omega^{d-3}} = r^2 \cos^2 \vartheta \mapsto [(r'+\bar{r}')/2] \cos^2 \vartheta = r^2 \cos^2 \vartheta^2, \nn\\
&& R^2 = r^2 + a^2 \cos^2 \vartheta.
\ee
The rotating metric is 
%\begin{widetext}
\be
 && ds^2 =  \frac{\Delta - a^2 \sin^2 \vartheta}{R^2} dt^2 
- \frac{ R^2}{\Delta }  \, dr^2 
- \Sigma \, d \vartheta^2  \nonumber \\
&& +  2 a \sin^2 \vartheta \left(1 - \frac{\Delta - a^2 \sin^2 \vartheta}{R^2} \right) dt \, d \phi 
\nonumber \\
 &&
 -  \, \sin ^2 \vartheta  \left[ R^2 + a^2 \sin^2 \vartheta \left(2 - \frac{\Delta - a^2 \sin^2\vartheta}{R^2}\right)   \right]    d \phi^2 \nonumber \\
&&  - r^2 \cos^2 \vartheta \, d \Omega^{d - 3}.
\label{ENCKN}
  \ee
Now $\Delta(r)$ is defined by the following relation
\be
&& \Delta = r^2 + a^2 - \frac{ 2 \mu(r)}{r^{d -4}}. \nn \\
&& 
\ee
The metric enjoys regularity at the origin as the previous four-dimensional solution. An important quantity for phenomenology is the temperature. To this purpose the surface gravity reads
%\begin{widetext}
\be
&&\kappa = \frac{r_+ ( d -2)}{2 ( r_+^2 + a^2 )} \nn \\
&& \left[ 1 + \left(\frac{d- 4}{d -2} \right) \frac{a^2}{r_+^2} 
- \frac{r_+^2 + a^2}{r_+ (d -2)} \frac{\gamma \left( \frac{d}{2}, \frac{r_+^2}{4 \ell^2} \right)'}{ 
\gamma \left( \frac{d}{2}, \frac{r_+^2}{4 \ell^2} \right) } \right]. 
\ee
%\end{widetext}
%
\subsection{Higher-dimensional regular Kerr-Newmann black hole}
The extension to higher dimension of the NCKN black hole follows the above line of reasoning. We start from
\be 
&& \hspace{-0.4cm} 
ds^2_{(d+1)} =G(r) dt^2 - 
\frac{dr^2}{G(r)}
%\nn \\
%&& 
- \underbrace{r^2}_{g_{\Omega^{2}}} (d \vartheta^2 + \sin^2 \vartheta \phi^2) \
\nn \\ && \hspace{0.8cm}
- \underbrace{ r^2 \cos^2 \vartheta}_{g_{\Omega^{d -3}}} d \Omega^{d-3},  \\
&& \hspace{-0.4cm} 
G(r)=\left(1 - \frac{2 \mu(r)}{r^{d-2}} +\frac{q(r)^2}{r^{2(d-2)}}\right) , \nn\\
&& \hspace{-0.4cm} 
\mu(r) = \frac{M}{M_{*}^{d -1} \Gamma(d/2)} \gamma(d/2; r^2/4 \theta) , \nn \\
&& \hspace{-0.4cm} 
q(r)^2 = (d-2)\frac{Q^2}{\pi^{d-2}}\left[F_d(r)+d_dr^{d-2}\gamma\left(\frac{d}{2}; \frac{r^2}{4 \theta}\right)\right] , \nn\\
&& \hspace{-0.4cm} 
F_d(r)=\gamma^2\left(\frac{d}{2}-1;\frac{r^2}{4 \theta}\right)-\frac{2^{\frac{8-3d}{2}}r^{d-2}}{(d-2)\theta^{\frac{d-2}{2}}}\gamma^2\left(\frac{d}{2}-1;\frac{r^2}{4 \theta}\right)  , \nn\\
&& \hspace{-0.4cm} 
d_d=\frac{2^{\frac{8-3d}{2}}}{d-2}\frac{1}{\theta^{\frac{d-2}{2}}}\frac{\Gamma\left(\frac{d}{2}-1\right)}{\Gamma\left(\frac{d}{2}\right)} , 
\ee
 and we apply our usual prescription for the complexification, i.e.
\be 
&& \hspace{-0.1cm} G(r) = 1 - \frac{2 \mu(r)}{r^2 r^{d-4}} + \frac{q(r)^2}{r^2 \, r^{2 ( d-3)}}  \nn \\
&&\mapsto 
1 - \frac{2 \mu[(r'+\bar{r}')/2]}{r' \bar{r}' \,  [(r' + \bar{r}' )/2]^{d-4}} +  
\frac{q[(r'+\bar{r}')/2]^2}{r' \bar{r}' \,  [(r' + \bar{r}' )/2]^{2(d-3)}}
 \nonumber \\
&& \hspace{0.1cm} 
=  1 -  \frac{ 2 \mu(r)}{R^2 \,  r^{d-4}} + \frac{q(r)^2}{R^2 \, r^{2 (d-3)}}, \nn \\
 && g_{\Omega^{2}} \mapsto r' \bar{r}' = R^2, \nn \\
 && g_{\Omega^{d-3}} = r^2 \cos^2 \vartheta \mapsto [(r'+\bar{r}')/2] \cos^2 \vartheta = r^2 \cos^2 \vartheta^2, \nn\\
&& R^2 = r^2 + a^2 \cos^2  \vartheta.
\ee
The rotating metric is (\ref{ENCKN})
with the new $\Delta(r)$ function 
\be
\Delta = r^2 + a^2 - \frac{ 2 \mu(r)}{r^{d -4}} + \frac{q(r)^2}{r^{2(d-3)}}.  
\ee
As a final note some comments are in order. The neutral metric (\ref{ENCKN}) approaches the Myers-Perry black hole with a single angular momentum  at large distances \cite{Myers:1986un} (see \cite{JNMP} for the derivation of the Myers-Perry black hole by employing the Newman-Janis procedure).
On the other hand we must have care about classical limits of the above charged solution. In Einstein gravity higher-dimensional spinning charged black holes have analytically been derived only in the slowly rotating case \cite{Aliev:2006yk}. For arbitrary values of the rotation parameter 
 the Einstein tensor coming from the Kerr-Shild decomposition or either from the Newman-Janis procedure fails to fit the higher-dimensional Maxwell spinning stress tensor. Only numerical solutions are known in the case of arbitrary rotation parameter \cite{kunz}.
As a consequence, since we have employed the Newman-Janis procedure and the higher-dimensional NCKN solution approaches the higher-dimensional Kerr-Newman line element at large distances we have the same problem as in the classical case. Therefore we can safely consider the solution only for small rotation parameter. As a result one obtains the higher-dimensional NCKN solution as a higher-dimensional NCRN solution \cite{extracharged} with additional first-order corrections in the rotation parameter.

%%%%%%%%%  COME PUOI VEDERE NEL PARAGRAFO SOPRA HO PROVATO A CALCOLARE
%ANCHE KERR NEWMANN MA POI HO SCOPERTO CHE IN REALTA' SI CONOSCIE SOLO UNA SOLUZIOENE APPROSSIMATA IN GR, QUINDI CREDO NON NE VALGA LA PENA.
\section{Conclusions}
\label{concl}
In this paper we have derived regular rotating charged and higher-dimensional black hole solution in the presence of a noncommutative geometry induced minimal length \cite{NCQFT,NCQFT2}. To get this result, we reviewed the Newman-Janis algorithm in order to include a wider class of spacetimes generated by nonvanishing stress tensors. Then we applied this method to obtain the rotating neutral black hole, which matches the known solution found in \cite{kerrr}. We have studied how noncommutative effects have smeared out the singularity, by calculating both the Ricci scalar and the Kretschmann invariant: they are both finite at the origin. Then we extended the analysis in \cite{kerrr} to the thermodynamic properties of this spinning solution, providing the black hole temperature. As a second step, we have applied our method to the case of a charged and spinning object, determining the noncommutative geometry inspired Kerr-Newman black hole. We studied the geometrical properties of the solution as well as the thermodynamic ones. In both frameworks the solution shows regularity. Finally we have determined the higher-dimensional extension of these metrics, providing their Hawking temperature.

With this paper, the program of noncommutative geometry inspired black holes, initiated in 2005 with the static, neutral solution, reached its completion.
As a final remark, there are some comments about future applications.
On the astrophysical side, charged black holes are often regarded as remote possibilities. On the ground of Newtonian gravity consideration it has been shown that the Coulomb repulsion would prevents the formation of black holes whose charge/mass ratio exceeds $10^{-18}$. This would mean that stellar black holes could have roughly $10^2$ C, a tiny charge compared for instance to the Earth charge which is about half a million C. However due to the rotation, an intense magnetic field might occur, for the presence of superficial currents on the hole. The phenomenology of these magnetic fields would in principle deserve further investigations.
On the other hand, the intrinsic quantum gravitational nature of these solutions at small length scales suggests that they could play a significant role in case of microscopic black hole phenomenology. Without entering the debate about the conjectured production of a mini black hole at the LHC, we recall that non-rotating higher-dimensional noncommutative geometry inspired black holes have been already the subject of investigations in TeV gravity phenomenology (analyses with Monte Carlo event generators can be found in \cite{QBH}). Even in the most pessimistic scenario of the absence of quantum gravity data at the Terascale, our new solutions can play a role at least in the context of primordial black holes. Indeed it is widely accepted that the early universe could have been populated by tiny black holes, which are
%The lack of their observation could be the explained by the fact that they are 
nowadays completely evaporated.
More specifically, the conventional scenario for the evaporation of a black hole consists of four distinct phases, the spin-down phase (loss of angular momentum), the balding phase (loss of long-range fields, i.e. hair, i.e. the electromagnetic charge), the Schwarzschild phase and the Planck phase. In light of the already determined NCBH solution this general evaporation scheme has been already reviewed. For instance the  final phase is turned into a new phase, called the SCRAM \footnote{The term SCRAM, probably the
acronym for ``Safety Control Rod Axe Man'' or ``Super Critical
Reactor Axe Man'' refers to an emergency shutdown of a
thermonuclear reactor. The term has been extended to cover
shutdowns of other complex operations or systems in an unstable
state, but it also has a standard meaning ``go away quickly,''
in particular, when we address children or animals. } phase, in which the black hole cools down to a zero temperature configuration instead of experiencing a runaway increase of the Hawking temperature \cite{NCthermo}. A further modification concerns the sequence of these phases since the discharging time depends on the number of extradimensions and on the kind of emission (brane, bulk-brane, bulk) governing both the Hawking and the Schwinger pair production mechanisms. Therefore the new NCKN solution is expected to provide the final answer about the black hole evolution from the very beginning to the very end, i.e. taking into account the yet unexplored spin-down mechanism too.

\begin{acknowledgments}
\noindent P.N. is supported by the Helmholtz International Center for FAIR within the
framework of the LOEWE program (Landesoffensive zur Entwicklung Wissenschaftlich-\"{O}konomischer Exzellenz) launched by the State of Hesse. P.N. would like to thank the Perimeter Institute for Theoretical Physics, Waterloo, ON, Canada for the kind hospitality during the period of work on this project. Research at Perimeter Institute is
supported by the Government of Canada through Industry Canada and by the Province of Ontario through the
Ministry of Research \& Innovation. The authors thank R. Balbinot, S. P. Drake and E. Spallucci for valuable suggestions and M. Sampaio for pointing out a typesetting inaccuracy in the manuscript.
\end{acknowledgments}

%\begin{widetext} 
\appendix
\section{Note on Gamma functions}
%\subsection{Note on Gamma functions}
%\noindent
The lower incomplete Gamma function is
\begin{equation}
\gamma\left( \frac{n}{2} ;  x \right)=\int_0^x\frac{dt}{t}\ t^{n/2}\ e^{t}.
\end{equation}
The upper incomplete Gamma function is
\begin{equation}
\Gamma\left( \frac{n}{2} ; x \right)=\int_x^\infty\frac{dt}{t}\ t^{n/2}\ e^{t}.
\end{equation}
The relation between the two is
\begin{equation}
\gamma\left(\ \frac{n}{2} ; x \right)=\Gamma\left( \frac{n}{2} \right)-\Gamma\left( \frac{n}{2} ;  x \right).
\end{equation}
From the relation
\begin{equation}
\Gamma\left( z+1 \right)= z \ \Gamma\left( z  \right)
\end{equation}
we find
\begin{equation}
\Gamma\left( \frac{3}{2} \right)= \Gamma\left( \frac{1}{2}+1 \right)=\ \frac{1}{2} \Gamma\left( \frac{1}{2}  \right)=\frac{1}{2} \sqrt{\pi}.
\end{equation}

%\vspace{0.5cm}
\section{Kretschmann invariant}
%\begin{widetext} 
The Kretschmann invariant for the spinning black hole ($Q = 0$), $r \geqslant 0$ and introducing 
the notation $\ell = \sqrt{\theta}$ is 
\vspace{-0.6cm}
\begin{widetext} 
\begin{eqnarray}
\label{KI}
&& K(r, \vartheta) = \frac{e^{-\frac{r^2}{2 \ell^2}} m^2 }
{8 \ell^{10} \pi 
   \left( \cos (2 \vartheta ) a^2+a^2+2 r^2\right)^6}
\Bigg(128 r^{16}+256 a^2 r^{14}+288 a^4 r^{12}+2048 \ell^4 r^{12}-1024 a^2 \ell^2 r^{12}  \nn \\ 
&& + 96 a^4   \cos (4 \vartheta ) r^{12}-2048 e^{\frac{r^2}{4 \ell^2}} \ell^5 \sqrt{\pi } r^{11}+160 a^6 r^{10}-1024 a^2 \ell^4 r^{10}-2304 a^4 \ell^2
   r^{10}+96 a^6 \cos (4 \vartheta ) r^{10} \nn \\
   &&-768 a^4 \ell^2 \cos (4 \vartheta ) r^{10}  +16 a^6 \cos (6 \vartheta ) r^{10}-8192
   e^{\frac{r^2}{4 \ell^2}} \ell^7 \sqrt{\pi } r^9+1024 a^2 e^{\frac{r^2}{4 \ell^2}} \ell^5 \sqrt{\pi } r^9+35 a^8 r^8-768 a^4 \ell^4
   r^8 \nn \\
   && -1920 a^6 \ell^2 r^8+28 a^8 \cos (4 \vartheta ) r^8 
    -256 a^4 \ell^4 \cos (4 \vartheta ) r^8-1152 a^6 \ell^2 \cos (4 \vartheta ) r^8+8 a^8
   \cos (6 \vartheta ) r^8-192 a^6 \ell^2 \cos (6 \vartheta ) r^8 \nn \\
   && + a^8 \cos (8 \vartheta ) r^8+1280 a^4 e^{\frac{r^2}{4 \ell^2}} \ell^5
   \sqrt{\pi } \cos (4 \vartheta ) r^7+53248 a^2 e^{\frac{r^2}{4 \ell^2}} \ell^7 \sqrt{\pi } r^7+3840 a^4 e^{\frac{r^2}{4 \ell^2}} \ell^5
   \sqrt{\pi } r^7+3200 a^6 \ell^4 r^6 \nn \\
&&   -560 a^8 \ell^2 r^6+1920 a^6 \ell^4 \cos (4 \vartheta ) r^6-448 a^8 \ell^2 \cos (4 \vartheta ) r^6+320
   a^6 \ell^4 \cos (6 \vartheta ) r^6-128 a^8 \ell^2 \cos (6 \vartheta ) r^6 \nn \\
   &&-16 a^8 \ell^2 \cos (8 \vartheta ) r^6+24576 e^{\frac{r^2}{2 \ell^2}}
   \ell^{10} \pi  r^6+5120 a^4 e^{\frac{r^2}{4 \ell^2}} \ell^7 \sqrt{\pi } \cos (4 \vartheta ) r^5+1152 a^6 e^{\frac{r^2}{4 \ell^2}} \ell^5
   \sqrt{\pi } \cos (4 \vartheta ) r^5 \nn \\
   && +192 a^6 e^{\frac{r^2}{4 \ell^2}} \ell^5 \sqrt{\pi } \cos (6 \vartheta ) r^5+15360 a^4
   e^{\frac{r^2}{4 \ell^2}} \ell^7 \sqrt{\pi } r^5+1920 a^6 e^{\frac{r^2}{4 \ell^2}} \ell^5 \sqrt{\pi } r^5+2240 a^8 \ell^4 r^4+1792 a^8 \ell^4
   \cos (4 \vartheta ) r^4 \nn \\
&&   +512 a^8 \ell^4 \cos (6 \vartheta ) r^4+64 a^8 \ell^4 \cos (8 \vartheta ) r^4-184320 a^2 e^{\frac{r^2}{2 \ell^2}}
   \ell^{10} \pi  r^4-13824 a^6 e^{\frac{r^2}{4 \ell^2}} \ell^7 \sqrt{\pi } \cos (4 \vartheta ) r^3 \nn  \\
&&   -2304 a^6 e^{\frac{r^2}{4 \ell^2}} \ell^7
   \sqrt{\pi } \cos (6 \vartheta ) r^3-23040 a^6 e^{\frac{r^2}{4 L^2}} \ell^7 \sqrt{\pi } r^3+46080 a^4 e^{\frac{r^2}{2 \ell^2}}
   \ell^{10} \pi  \cos (4 \vartheta ) r^2+138240 a^4 e^{\frac{r^2}{2 L^2}} \ell^{10} \pi  r^2 \nn \\
&&   -3072 e^{\frac{r^2}{2 \ell^2}} \ell^{10}
   \big(\cos (6 \vartheta ) a^6+10 a^6-180 r^2 a^4+6 \left(a^2-10 r^2\right) \cos (4 \vartheta ) a^4+240 r^4 a^2+15 \big(a^4-16
   r^2 a^2+16 r^4\big)  \nn \\
&&  \cos (2 \vartheta ) a^2 -32 r^6 \big) \Gamma \left(\frac{3}{2},\frac{r^2}{4 \ell^2}\right)^2+8 a^2 \Big(7
   r^4 \left(r^2-8 \ell^2\right)^2 a^6-30 \Big(48 e^{\frac{r^2}{2 \ell^2}} \pi  \ell^{10}+12 e^{\frac{r^2}{4 \ell^2}} \sqrt{\pi } r^3
   \left(12 \ell^2-r^2\right) \ell^5 \nn \\
  && -r^6 \left(20 \ell^4-12 r^2 \ell^2+r^4\right) \Big) a^4+16 r^2 \left(1440 e^{\frac{r^2}{2 \ell^2}} \pi 
   \ell^{10}+40 e^{\frac{r^2}{4 \ell^2}} \sqrt{\pi } r^3 \left(4 \ell^2+r^2\right) \ell^5-8 r^6 \ell^4-24 r^8 \ell^2+3 r^{10}\right) a^2 \nn  \\
 %  \ee
  % \be
   %
   &&+32 r^4
   \left(-720 e^{\frac{r^2}{2 \ell^2}} \pi  \ell^{10}+4 e^{\frac{r^2}{4 \ell^2}} \sqrt{\pi } r^3 \left(52 \ell^2+r^2\right) \ell^5-4 r^6
   \ell^4-4 r^8 \ell^2+r^{10}\right) \Big) \cos (2 \vartheta )-4608 a^6 e^{\frac{r^2}{2 \ell^2}} \ell^{10} \pi  \cos (4 \vartheta ) \nn \\
&&   -768 a^6
   e^{\frac{r^2}{2 \ell^2}} \ell^{10} \pi  \cos (6 \vartheta )-128 e^{\frac{r^2}{4 \ell^2}} \ell^5 \Big(-32 r^{11}+16 a^2 r^9-128 \ell^2
   r^9+60 a^4 r^7+832 a^2 \ell^2 r^7+768 e^{\frac{r^2}{4 \ell^2}} \ell^5 \sqrt{\pi } r^6 \nn \\
   &&+30 a^6 r^5+240 a^4 \ell^2 r^5+3 a^6 \cos (6
   \vartheta ) r^5-5760 a^2 e^{\frac{r^2}{4 \ell^2}} \ell^5 \sqrt{\pi } r^4-360 a^6 \ell^2 r^3-36 a^6 \ell^2 \cos (6 \vartheta ) r^3+4320 a^4
   e^{\frac{r^2}{4 \ell^2}} \ell^5 \sqrt{\pi } r^2 \nn \\
   && -a^2 \Big(45 \left(8 e^{\frac{r^2}{4 \ell^2}} \sqrt{\pi } \ell^5+12 r^3 \ell^2-r^5\right)
   a^4-80 r^2 \left(72 e^{\frac{r^2}{4 \ell^2}} \sqrt{\pi } \ell^5+4 r^3 \ell^2+r^5\right) a^2 
   -16 r^4 (-360 e^{\frac{r^2}{4 \ell^2}}
   \sqrt{\pi } \ell^5+52 r^3 \ell^2
   \nn \\
  && +r^5 ) \Big) \cos (2 \vartheta )-2 a^4 \Big(9 a^2 \left(8 e^{\frac{r^2}{4 \ell^2}} \sqrt{\pi }
   \ell^5+12 r^3 \ell^2-r^5\right)  -10 r^2 \left(72 e^{\frac{r^2}{4 \ell^2}} \sqrt{\pi } \ell^5+4 r^3 \ell^2+r^5\right)\Big) \cos (4 \vartheta
   )  \nn \\
   && -24 a^6 e^{\frac{r^2}{4 \ell^2}} \ell^5 \sqrt{\pi } \cos (6 \vartheta )-240 a^6 e^{\frac{r^2}{4 \ell^2}} \ell^5 \sqrt{\pi }\Big)
   \Gamma \left(\frac{3}{2},\frac{r^2}{4 \ell^2}\right) 
  -7680 a^6 e^{\frac{r^2}{2 \ell^2}} \ell^{10} \pi \Bigg).  \end{eqnarray}
%\end{widetext}
%
\vspace{3cm}
\section{Stress-energy tensor components}
%\begin{widetext}
\be
&&T^t_t = \frac{1}{16 \pi  \left(a^2 \cos (2 \vartheta
   )+a^2+2 r^2\right)^3}
    \Big[ 2 a^4 r \cos (4 \vartheta ) m''(r)-2 a^4 r m''(r)-a^4 \cos (4 \vartheta ) [q(r)^2]''+a^4 [q(r)^2]''  \\
    &&
    +8 a^2 r^3
   \cos (2 \vartheta ) m''(r)-8 a^2 r^3 m''(r)-4 a^2 r^2 \cos (2 \vartheta ) [q(r)^2]'' + 4 a^2 r^2 [q(r)^2]'' + 8 a^2 r
   \cos (2 \vartheta ) [q(r)^2]'  \nn \\
   && - 24 a^2 r [q(r)^2]'
   -8 [q(r)^2] \left(a^2 \cos (2 \vartheta )-3 a^2-2 r^2\right)+4 m'(r)
   \left(a^4 \cos (4 \vartheta )-a^4+8 a^2 r^2+8 r^4\right)-16 r^3 [q(r)^2]'  \Big] , \nn
   \ee
   \be
   T^r_r = \frac{r \left(2 r m'(r)-[q(r)^2]' \right)+[q(r)^2]}{2 \pi  \left(a^2 \cos (2 \vartheta )+a^2+2 r^2\right)^2} ,
   \ee
   \be
  && T^{\vartheta}_{\vartheta} = \frac{1}{8 \pi  \left(a^2 \cos (2 \vartheta )+a^2+2
   r^2\right)^2}
   \Big[2 a^2 r \cos (2 \vartheta ) m''(r)+2 a^2 r m''(r)+8 a^2 \cos ^2(\vartheta ) m'(r)-a^2 \cos (2 \vartheta )
   [q(r)^2]''  \nn \\
   && - a^2 [q(r)^2]'' +4 r^3 m''(r)-2 r^2 [q(r)^2]''+4 r [q(r)^2]'(r)-4 q(r)^2 \Big] ,
   \ee
   \be
   && T^{\phi}_{\phi} = 
  \frac{1}{4 \pi  \left(a^2 \cos (2 \vartheta )+a^2+2 r^2\right)^3}
 \Big[ 2 a^4 r \cos (2 \vartheta ) m''(r)+2 a^4 r m''(r)-a^4 \cos (2 \vartheta ) [q(r)^2]''-a^4 [q(r)^2]''   \\
 &&
 +2 a^2 r^3
   \cos (2 \vartheta ) m''(r)+6 a^2 r^3 m''(r)+4 a^2 m'(r) \left(\left(a^2+2 r^2\right) \cos (2 \vartheta
   )+a^2\right)-a^2 r^2 \cos (2 \vartheta ) [q(r)^2]''
   -3 a^2 r^2 [q(r)^2]'' \nn \\
   &&
   -2 a^2 r \cos (2 \vartheta ) [q(r)^2]'+6 a^2 r
   [q(r)^2]'+2 [q(r)^2] \left(a^2 \cos (2 \vartheta )-3 a^2-2 r^2\right)+4 r^5 m''(r)-2 r^4 [q(r)^2]''+4 r^3 [q(r)^2]' \Big] ,
  \nn
   \ee 
\be
   && T^{\phi}_t = - \frac{a}{4 \pi  \left( a^2 \cos (2 \vartheta )+a^2+2 r^2 \right)^3}
  \Big[  2 a^2 r \cos (2 \vartheta ) m''(r)+2 a^2 r m''(r)+4 m'(r) \left(a^2 \cos (2 \vartheta )+a^2-2
   r^2\right) \nn \\
   &&
   -a^2 \cos (2 \vartheta ) [q(r)^2]''-a^2 [q(r)^2]''+4 r^3 m''(r)-2 r^2 [q(r)^2]''+8 r [q(r)^2]'-8
   [q(r)^2]  \Big] .
   \ee
   %
 %  \be
  % && T^t_{\phi} = \frac{a \left(a^2+r^2\right) \sin ^2(\vartheta )}{4 \pi  \left(a^2 \cos (2 \vartheta )+a^2+2 r^2\right)^3}
  %   \Big[2 a^2 r \cos (2 \vartheta ) m''(r)+2 a^2 r m''(r)+4
 %  m'(r) \left(a^2 \cos (2 \vartheta )+a^2-2 r^2\right) \nn \\
 %  &&
 %  -a^2 \cos (2 \vartheta ) [q(r)^2]''-a^2 [q(r)^2]''+4 r^3
%   m''(r)-2 r^2 [q(r)^2]''+8 r [q(r)^2]'-8 [q(r)^2]   \Big]
   %
   %   %
%
%
%
%\ee
\end{widetext}
%%%%%

\end{document}